# Dynamic Systems Model for Filamentary Mem-Resistors

Blaise Mouttet

*Abstract*— **A dynamic systems model is proposed describing memory resistors which include a filament conductive bridge. In this model the system state is defined by both a dynamic tunneling barrier (associated with the filament-electrode gap) and a dynamic Schottky barrier (associated with the electron depletion width surrounding the filament-electrode gap). A general model is formulated which may be applicable to many different forms of memory resistor materials. The frequency response of the model is briefly discussed.**

## I. INTRODUCTION

A dynamic systems model was recently proposed for ionic mem-resistors based on harmonic oscillation of either electronic or ionic depletion widths in metal-semiconductor junctions [1]. This model was made under the assumption that the depletion width was uniform across the area of the junction. However, this assumption is not justified for certain metal oxides [2] and ion-doped chalcogenides [3] which include localized conducting filaments. In the 1990's and 2000's research and development performed by scientists of Axon Technology and Micron Technology have demonstrated the existence of such conducting filaments as important to solid electrolyte memory cells [4]. The filaments may be formed in various memory resistor materials by different mechanisms such as accumulation of ions or vacancies in non-uniform electric fields or via electrochemical reactions at an active electrode [5].

The present article expands the harmonic mem-resistor model of [1] to incorporate the effects of a filament in an ionic junction.

## II. FILAMENT DYNAMICS IN THIN FILMS

Following the approach of [1] this model expresses the dynamic equation in terms of Newton's 2$^{nd}$ law of motion relating the acceleration $d^2x_f/dt^2$ of the tip of a filament having effective mass $m_f$ to the sum of the forces $F_i$ acting on it.

$$m_f \frac{d^2 x_f}{dt^2} = \sum_i F_i \quad (1)$$

When an external electric field is applied to a filament in a thin film sandwiched between two electrodes there are three principle forces which act on the filament tip. The first force ($F_c$) is due to collisions of the tip with the surrounding media as the tip-electrode gap varies. The product of this force and the average time between collisions $\tau_c$ can be equated to the change in the tip momentum.

$$F_c \tau_f = -m_f \frac{dx_f}{dt} \quad (2)$$

The second force ($F_r$) is due to the internal electric field ($E_r$) seen by ions attached to the filament tip. Fig. 1 provides an approximate picture of a conducting filament tip as it approaches an electrode. As the gap between the tip and the electrode approaches the width of a tunneling gap charged ions or vacancies will either become attached to the filament tip or swept away from the gap region due to electrostatic forces. This will result in the depletion of ions within the tunneling gap. As a result of this ion depletion the 2DEG normally found in the metal side of a Schottky junction will be neutralized. The charge neutrality of the interface allows the use of the method of image charges to be used to calculate the internal electric field in the region between the tip and the electrode. The exact calculation of this internal field would require knowledge of the geometry of the tip. However, an approximate solution can be determined using the mean of the equilibrium positions of the collective ions denoted by $x_{f0}$. In this case an application of Gauss's Law produces

$$F_r = zeE_r = -\frac{(ze)^2 n_f}{A_f x_f(t)\varepsilon_r \varepsilon_0}(x_f(t) - x_{f0}) \quad (3)$$

where $\varepsilon_0$ is the vacuum permittivity, $\varepsilon_r$ is the relative permittivity, $e$ is the unit charge, $z$ is the valence of the ions, $n_f$ is the number of ions on the filament tip, $A_f$ is the cross-sectional area of the filament tip, and $x_f(t)$ is the dynamic tunnel gap.

The third force $F_a$ is related to the externally applied voltage bias by a proportionality constant $K$ determined by the tip geometry.

$$F_a = -KzeE_a = KzeV_a/x_f(t) \quad (4)$$

It is notable that this force is expected to be 180 degrees out of phase with the force seen by the ions in the Schottky region since as ions are repelled from the gap by the applied potential they would be swept away from the tunnel gap which would cause the gap to decrease. On the other if the potential is such that it attracts ions to the junction the resultant electrostatic forces would cause the tunnel gap to increase.

Combining (1)-(4) produces:

$$m_f \frac{d^2 x_f}{dt^2} + \frac{m_f}{\tau_f}\frac{dx_f}{dt} + \frac{(ze)^2 n_f}{A_f x_f(t)\varepsilon_r \varepsilon_0}(x_f(t) - x_{f0}) = \frac{KzeV_a(t)}{x_f(t)} \quad (5)$$

A simplified version of (5) may be developed in the case where the maximum deflection of the tip $\Delta x_f(t)$ is small compared to the equilibrium position $x_{f0}$.

$$\Delta x_f(t) = x_f(t) - x_{f0} \quad (6)$$

$$\Delta x_f(t) \ll x_{f0} \quad (7)$$

$$\frac{d^2 \Delta x_f(t)}{dt^2} + \frac{1}{\tau_f}\frac{d\Delta x_f(t)}{dt} + \frac{(ze)^2 n_f}{m_f x_{f0} A_f \varepsilon_r \varepsilon_0}\Delta x_f(t) \approx \frac{KzeV_a(t)}{m_f x_{f0}} \quad (8)$$

As in [1] we arrive at a tractable form in the form of the familiar driven damped harmonic oscillator differential equation.

$$\frac{d^2 x}{dt^2} + 2\zeta\omega_0 \frac{dx}{dt} + \omega_0^2 x = F(t) \quad (9)$$

### III. COUPLING FILAMENT AND ELECTRONIC DYNAMICS

The analysis of [1] for dynamic tunneling junctions can now be repeated in the case of the tunneling gap. When the applied voltage to this system is zero ($V_a(t)=0$) the magnitude of the tunneling energy barrier $\Phi_{B0}(t)$ is the product of the electric field $E_0$ in the gap and the ion depletion width $x_f(t)$.

$$\Phi_{B0}(t) = E_0 x_f(t) \quad (10)$$

$E_0$ may be approximated as

$$E_0 = \frac{ze n_f}{A_f \varepsilon_r \varepsilon_0} \quad (11)$$

and $\Phi_{B0}(t)$ is

$$\Phi_{B0}(t) = \frac{ze n_f}{A_f \varepsilon_r \varepsilon_0} x_f(t) \quad (12)$$

At zero voltage some tunneling between the metal and ionic region may occur due to the thermal energy of the electrons. At equilibrium the tunneling current density $J_{T0}$ from the metal to the ionic region should balance the tunneling current density from the ionic region to the metal and can be calculated using the tunneling current equation as referenced in [1]. Note that $x_f(t)$ is time-dependent but is a constant for purposes of the integration with respect to x.

$$J_{T0}(t) = C_0 \exp\left(\frac{-\sqrt{8m_e}}{h/2\pi}\int_0^{x_f(t)}\sqrt{\Phi_{B0}(t)}dx\right) = C_0 \exp\left(\frac{-\sqrt{8m_e}}{h/2\pi}\sqrt{\frac{ze n_f}{A_f \varepsilon_r \varepsilon_0} x_f^3(t)}\right) \quad (13)$$

As a positive voltage is applied to the left electrode the height of the barrier decreases so that

$$\Phi_{Bv}(t) = E_0 x_f(t) - V_a(t) \quad (14)$$

and the tunneling current density is now calculated as

$$J_{Tv}(t) = C_0 \exp\left(\frac{-\sqrt{8m_e}}{h/2\pi}\int_0^{x_{d0}(t)}\sqrt{\Phi_{Bv}(t)}dx\right) =$$
$$C_0 \exp\left(\frac{-\sqrt{8m_e}}{h/2\pi}\sqrt{\frac{zen_f}{A_f\varepsilon_r\varepsilon_0}x_f^3(t) - V_a(t)x_f^2(t)}\right) \quad (15)$$

The net increase of current density from equilibrium is

$$J_T(t) = J_{Tv}(x_f(t)) - J_{T0}(x_f(t)) \quad (16)$$

TABLE 1 summarizes the equations for the tunneling filament in addition to the dynamic Schottky and capacitance components discussed in [1]. The total current density $J$ will of course be a weighted average of the tunneling $J_T$, Schottky $J_S$, and capacitance $J_C$ current densities in accordance with

$$J(t) = \frac{[J_S(t)+J_C(t)](A-A_t)+J_T(t)A_t}{A} \quad (17)$$

where $A$ is the total electrode area and $A_t$ is the cross-sectional area of the filament.

*c) Frequency Response*

It is expected that capacitance effects dominate the frequency response of the filamentary mem-resistor fabricated as memory cells. However, for experimental cases a scanning tunneling microscope may be used as the electrode on which the filament grows so that $A \approx A_t$. In this case a sinusoidal voltage may be applied to the dynamic tunneling junction

$$V_a(t) = V_0 \sin(\omega t) \quad (18)$$

and the steady-state solution to (8) takes the form

$$\Delta x_f(t) = \Delta X_{f0} \sin(\omega t + \varphi_0) \quad (19)$$

$$\Delta X_{f0} = \frac{KzeV_0}{m_f x_{f0}\sqrt{(\omega/\tau_f)^2 + \left(\omega^2 - \frac{(ze)^2 n_f}{m_f x_{f0} A_f \varepsilon_r \varepsilon_0}\right)^2}} \quad (20)$$

$$\varphi_0 = \tan^{-1}\frac{\omega}{\left(\omega^2 - \frac{(ze)^2 n_f}{m_f x_{f0} A_f \varepsilon_r \varepsilon_0}\right)\tau_f} \quad (21)$$

Similarly to the situation noted in [1] at resonance $\varphi_0 = 90$ degrees and the dynamic behavior of the tunneling width is 90 degrees out of phase with the applied voltage. As a result a zero-crossing hysteresis curve will develop in the current vs. voltage curve. As the input signal frequency increases or decreases sufficiently from the resonance frequency the phase shift $\varphi_0$ will go to zero and the hysteresis effect will disappear.

## IV. CONCLUSION

This paper has provided a model of resistance switching of filamentary memory resistors. It is hoped that the equations summarized in TABLE 1 will be of assistance to further development of ReRAM. It is also hoped that they will assist to further develop my patented inventions involving mem-resistor crossbars used in signal processing circuits and robotic control systems [6].

*TABLE 1 Summary of Equations for Filamentary Memory Resistor*

## Schottky and capacitive components

$$J_S(t) = J_{S0}(t)\left[\exp\left(\frac{zeV_a(t)}{kT}\right) - 1\right]$$

$$J_{S0}(t) = \frac{(ze)^2 D_n N_c}{kT}\left[\frac{2ze\left(\frac{ze}{\epsilon_r\epsilon_0}(N_d x_{d0}(x_{d0}+2\Delta x_d(t)) - V_a(t))N_d x_{d0}\right)}{\epsilon_r\epsilon_0(x_{d0}+2\Delta x_d(t))}\right]^{1/2} \exp\left(\frac{-ze\phi_B}{kT}\right)$$

$$J_c(t) = \frac{d[C(t)V_a(t)]}{dt} = \frac{d}{dt}\left[\frac{\epsilon_r\epsilon_0}{x_{d0}+2\Delta x_d(t)}V_a(t)\right]$$

$$\frac{d^2\Delta x_d(t)}{dt^2} + \frac{1}{\tau_c}\frac{d\Delta x_d(t)}{dt} + \frac{(ze)^2 N_d}{m_{ion}\epsilon_r\epsilon_0}\Delta x_d(t) = -\left(\frac{a^2 v ze}{\tau_c kT}\right)\exp\left(\frac{-W_a}{kT}\right)\frac{V_a(t)}{x_{d0}}$$

## Tunneling component

$$J_T(t) = C_0\left(\exp\left(\frac{-\sqrt{8m_e}}{h/2\pi}\sqrt{\frac{zen_f}{A_f\varepsilon_r\varepsilon_0}x_f^3(t) - V_a(t)x_f^2(t)}\right) - \exp\left(\frac{-\sqrt{8m_e}}{h/2\pi}\sqrt{\frac{zen_f}{A_f\varepsilon_r\varepsilon_0}x_f^3(t)}\right)\right)$$

$$\frac{d^2\Delta x_f(t)}{dt^2} + \frac{1}{\tau_f}\frac{d\Delta x_f(t)}{dt} + \frac{(ze)^2 n_f}{m_f x_{f0} A_f \varepsilon_r \varepsilon_0}\Delta x_f(t) = \frac{kze}{m_f}\frac{V_a(t)}{x_{f0}}$$

$$\Delta x_f(t) = x_f(t) - x_{f0}$$

## Total Dynamic Current Density

$$J(t) = \frac{[J_S(t) + J_C(t)](A - A_f) + J_T(t)A_f}{A}$$

June 25, 2011 (ver.3)

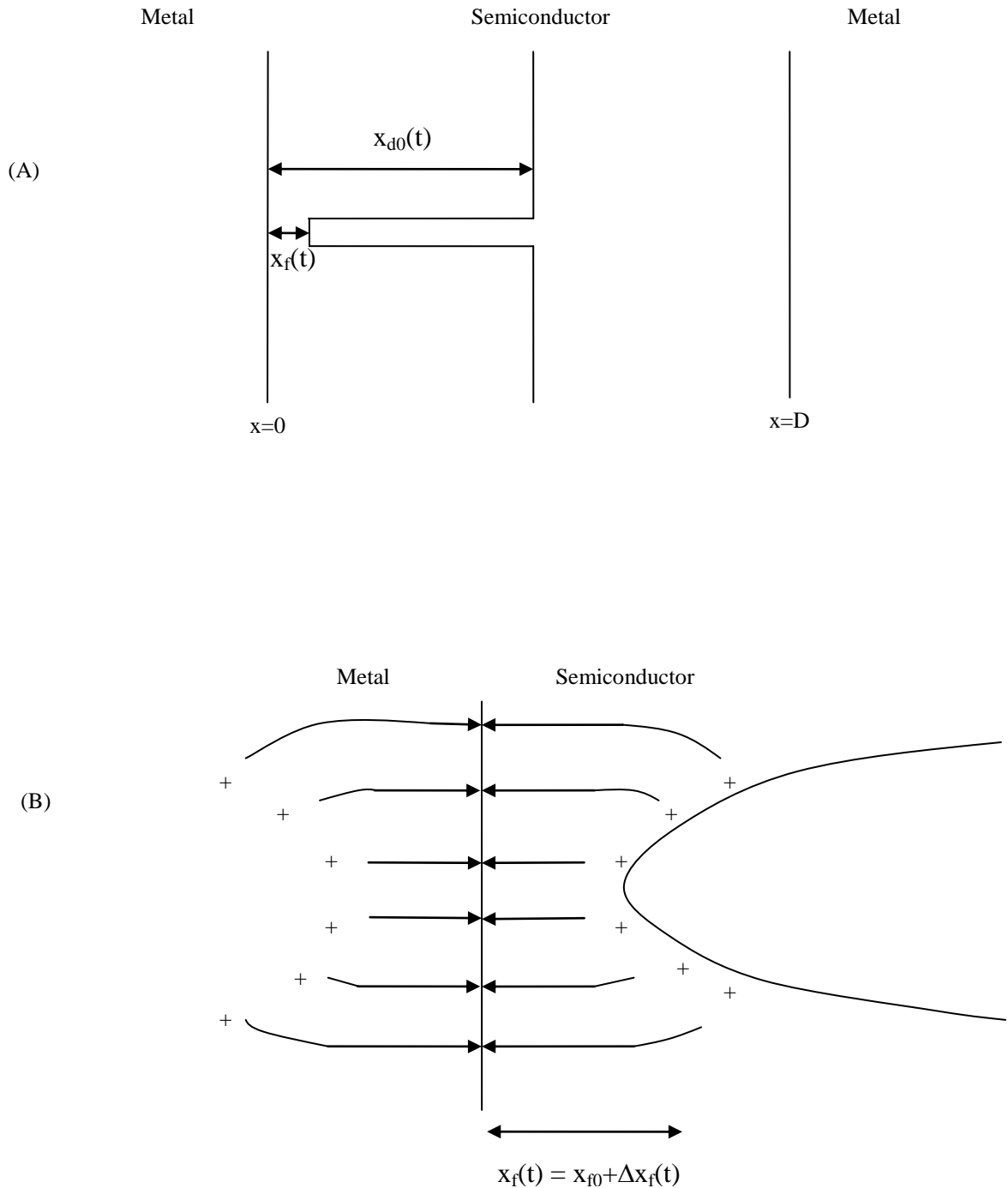

Fig. 1
(A) Illustration of a dynamic electron depletion width $x_{d0}(t)$ surrounding a filament and an ionic depletion width $x_f(t)$ between the filament tip and electrode in a metal-semiconductor-metal cell.
(B) A close up view of the ionic depletion width $x_f(t)$ between the filament tip and the left electrode illustrating mirror charges used in determining the electrostatic field.

June 25, 2011 (ver.3)